\documentclass[a4paper,11pt]{article}
\usepackage{pos}

\title{Relativistic MHD simulations of merging and collapsing stars and effects on GRB transients}
\ShortTitle{GR MHD simulations of mergers and collapsars}

\author*[a]{Agnieszka Janiuk}
\author[a]{Gerardo Urrutia}
\author[a]{Joseph Saji}
\author[a]{Piotr Plonka}

\affiliation[a]{Center for Theoretical Physics,\\
  Al. Lotnikow 32/46, 02-668, Warsaw, Poland}

\emailAdd{agnes@cft.edu.pl}
\abstract{Compact binary mergers and the collapse of massive stars can produce intense transients observable across high-energy wavelengths. Events such as gamma-ray bursts and kilonova emissions
are often accompanied by gravitational wave detections, making them crucial sources for multimessenger astrophysics. To explore these phenomena theoretically, state-of-the-art approaches of General Relativistic magnetohydrodynamic simulations are used. We present recent findings
from our simulations, and discuss observational consequences of the stellar/post-merger environment
on the gamma ray burst prompt emission properties.}

\FullConference{
}

\begin{document}
\maketitle

\section{Introduction}

Long and short gamma ray bursts (GRBs) are subclasses of transient phenomena associated with emission from relativistic jets launched by a central engine.
These jets produce gamma rays when they expand with almost speed of light velocities and their radiation is beamed towards the observer's line of sight.
In case of long GRBs, the progenitor, a very massive star (more than 20 $M_{\odot}$), can affect the propagation of jets that need to break out through the stellar envelope.
An accretion disk formed after gravitational collapse of the stellar core, must accelerate the jet to a sufficiently high speed and provide enough energy for the jet to break out.

In case of short GRBs, a similar type of central engine, namely accreting black hole, is often invoked. Here the jet energetics and speed may also be similar, even though there is no stellar envelope on the way. Still, the material launched before and during the compact binary merger provides environment that is interacting with the jet and that may affect its properties.
%, such a collimation angle.

Magnetic fields play an important role in the process of jet launching and magnetized accretion onto black hole provides sufficiently large energy to the jets. Specifically, in the magnetically arrested mode of accretion (MAD state) \cite{tchekhovskoy2011} the jet efficiency %(i.e. its power, 
$\eta=L_{\rm jet}/\dot M c^{2} $ %with respect to the accretion energy, 
is large. 
The magnetic fields anchored in the accretion disk penetrate ergosphere and mediate extraction of the black hole's rotational energy. The poloidal field accumulated in the BH horizon region builds up a temporary barrier, which acts as a channeling mechanism for the plasma to reach the horizon \cite{knalew}. 
Variability of this state may manifest later in the intermittency of jet launching process, and ultimately lead to a highly variable prompt emission of GRBs \cite{lloyd2016}. It was found to be correlated with the fastest growing mode of the magneto-rotational instability in the accretion disk \cite{sapountzis2019}.

%relativistic jets

%MAD mode of accretion

%Jet launching and its variability

\section{Our methodology and numerical simulations}

Our methodology adopted to study jet launching and properties of jets in GRBs from mergers and collapsars is based on solving the standard equations of magnetohydrodynamics in General Relativity. We employ numerical simulations based on the conservative scheme for ideal MHD and follow the evolution of conserved variables computed from the stress-energy tensor. The equations stand for the energy-momentum and mass-energy conservation, and are supplemented by Maxwell equations for the evolution of electromagnetic fields. Thermodynamics of the gas is treated by the equation of state (EOS) which in the simplest form can be taken as adiabatic relation between internal energy and pressure. 

In more elaborate simulations, we focus on the microphysics of the accretion flow, and its pressure and temperature changes due to chemical evolution. In particular, post-merger accretion disks are very neutron rich, and their composition evolves due to nuclear reactions (electron-positron capture on nucleons, pair annihilation, synthesis of Helium nuclei). The nuclear heating, and also cooling by neutrinos emitted in these reactions, must be incorporated in the MHD scheme where it acts as an extra source term in the energy equation \cite{Janiuk2025}.

Finally, the flow in collapsing stars, even though less dense than in mergers, is quite massive, and hence subject to the self-gravity force. Below, we present results of our calculations of GRB engines and jets assuming: (i) adiabatic EOS in Sect. \ref{sect:jets}, (ii) nuclear EOS in Sect. \ref{sect:nucleo} and Sect. \ref{sect:winds}, and (iii) self-gravity forces
in Sect. \ref{sect:collapsar}.

\section{Jet profiles and energetics}
\label{sect:jets}

The jets arising from our GR MHD simulations are highly non-uniform and variable, see \cite{janiuk2021}. As shown in \cite{saji2025}, the jet energetics varies across the polar angle, and its profile depends on the engine parameters: mass of the disk and black hole spin. The highest terminal Lorentz factors around $\Gamma\sim 500$ were obtained in the simulation with a large BH spin $a=0.9$, while for smaller spins of $a\sim 0.6-0.8$ the jets were not as energetic  and reached 
$\Gamma \sim 100-200$. The highly energetic jets exhibited concentration of Lorentz factor close to the axis, and for less energetic jets this distribution was spreading more towards jet edges. 
We obtained opening angles of the jet, measured in the simulations as the region of polar angles where $75\%$ of the total energy is contained, to be between $8-25^{\circ}$. These angles match several known short GRBs, where the opening angles have been estimated from the afterglow observations. In particular, the source GRB 090510 was examined in our work taking into account the constraints for its energetics from isotropic total energy measurements, $E_{\rm iso} \sim 9.97\times 10^{52}$~erg, and opening angle estimation $\Theta_{\rm jet}\sim 10^{\circ}$. A plausible model representing this source involved a BNS post-merger system with a 3 $M_{\odot}$ spinning black hole and a $0.07 M_{\odot}$ accretion disk.

  \begin{figure}
  \centering
    \includegraphics[width=0.9\textwidth]{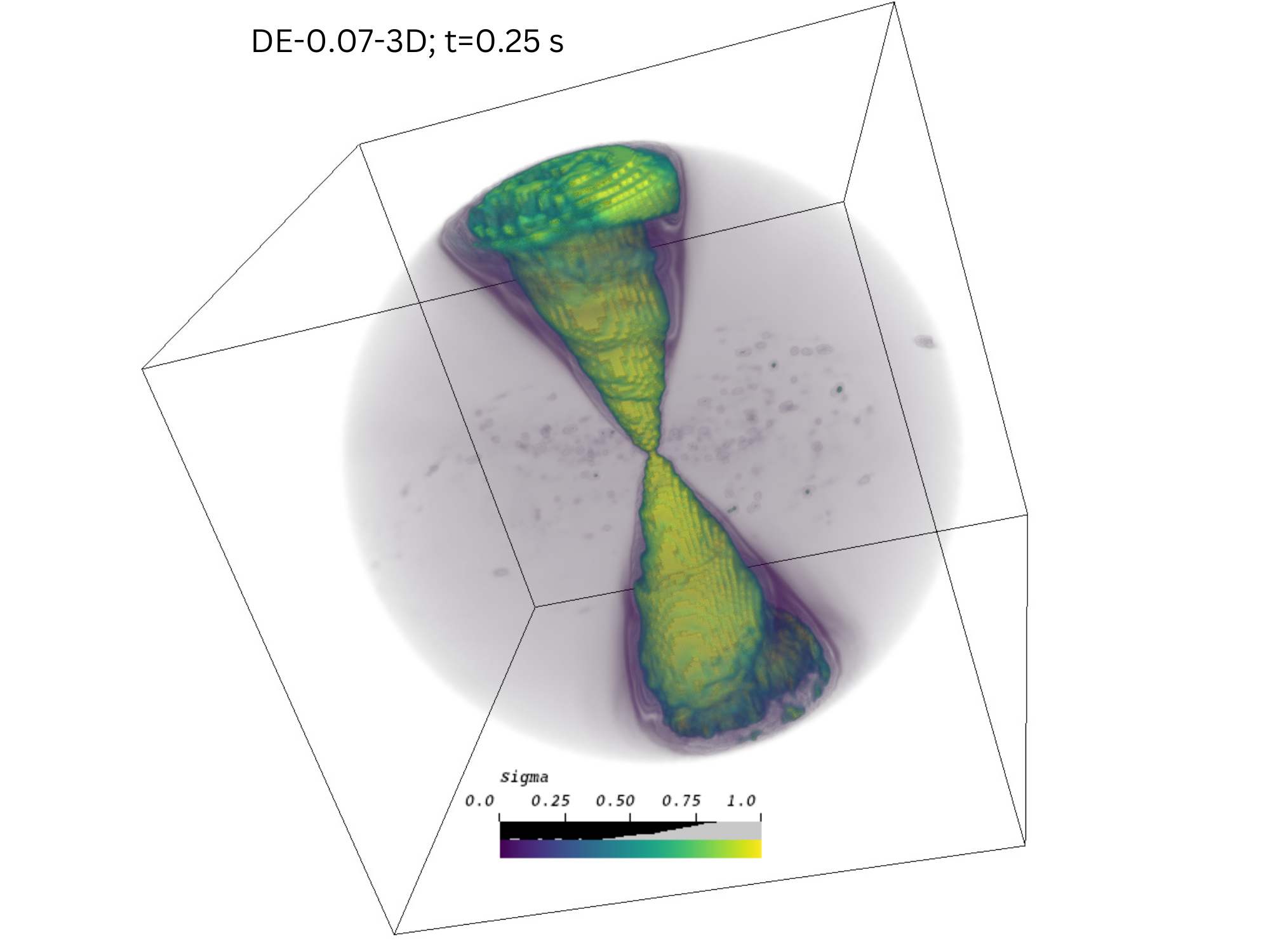}
\caption{Jet structure of an evolved 3D model assuming 
%jet interaction with 
presence of Dynamical Ejecta.
%in our sample, HD-0.10-3D. 
The parameter represented by color scale is jet magnetization $\sigma$, and the bounding box is $500 r_{g}$}.
\label{fig:jet_profile}
\end{figure}

In Figure \ref{fig:jet_profile} we present profile of the jet representing its total (thermal plus Poynting) energy distribution, taken at evolved time from our 3D simulation.
In this particular model, we adopted an additional medium, so called Dynamical Ejecta, that represent material expelled during or before the merger.% of two compact objects. 
Our aim was to test whether these ejecta can act as additional collimation mechanism for the jet, or if they can affect its variability properties.
No such effect was confirmed in the simulations, possibly because the jet was too much energetic ($E_{jet}\sim 1.2 \times 10^{52}$ erg). A weaker jet is expected to be affected by DE in a more effective way (H. Hamidani, private communication).

\section{Accretion disk winds and nucleosynthesis}
\label{sect:nucleo}

 During the dynamical evolution of black hole accretion disks in GRB engines, magnetically driven winds emerge and are accelerated to mildly-relativistic velocities. It has been shown in our previous work \cite{janiuk2019} that these winds can be efficient sites of r-process nucleosynthesis and eject copious amounts of heavy elements in the unbound ejecta. We model the BH-disk system by means of GR MHD simulations with a composition dependent, 3-parameter EOS, with pressure and internal energy dependent on plasma parameters: rest mass density, temperature and electron fraction. The electron fraction is evolving dynamically because the lepton number density is changing due to weak interactions. The disk is cooled by neutrinos, and energy change per unit volume due to neutrino emission provides a non-zero source term to the energy equation in our GR MHD system, that must be solved with specific recovery schemes (see e.g. \cite{siegel2018}).

  \begin{figure}
  \centering
    \includegraphics[width=0.9\textwidth]{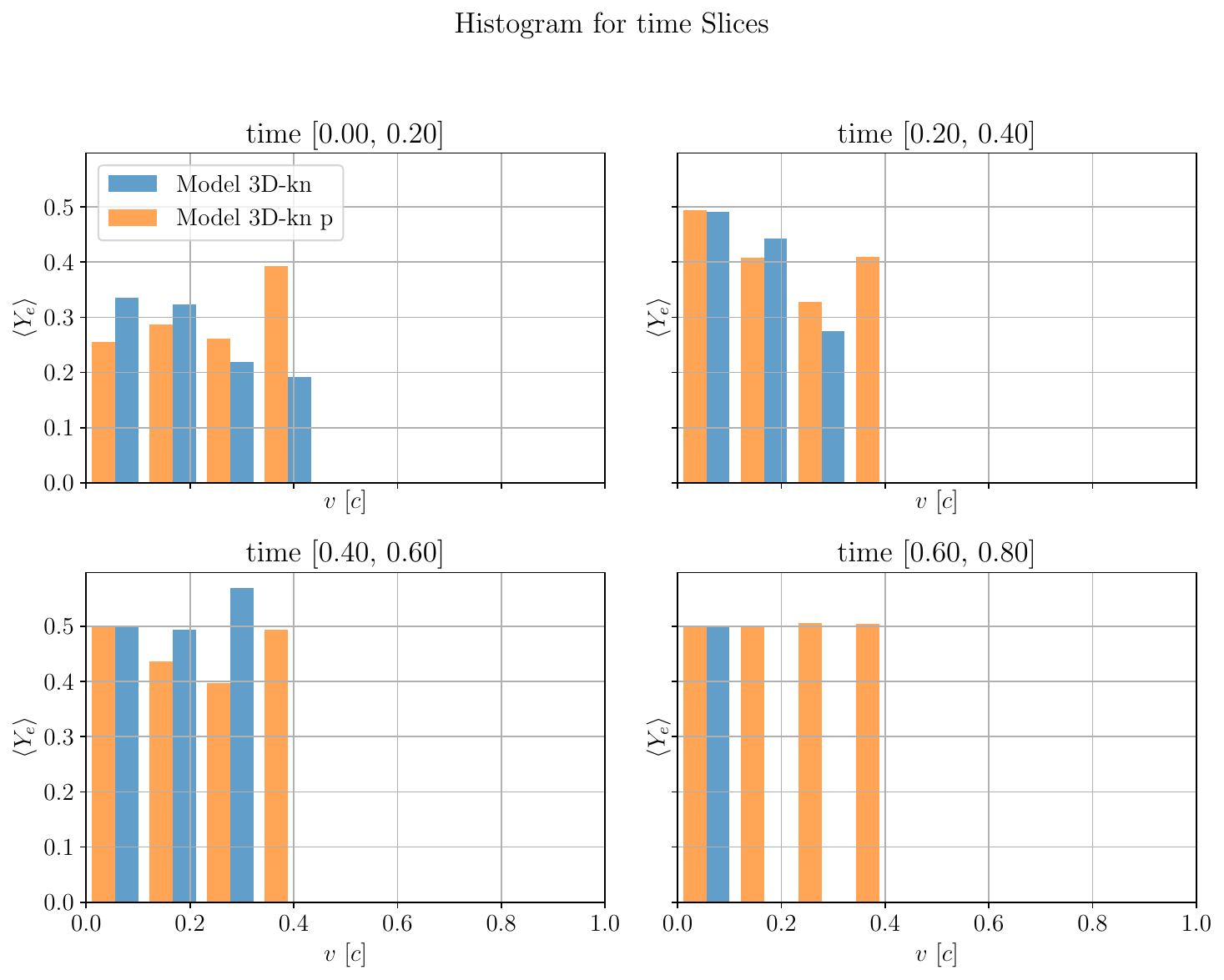}
    \caption{Histograms of electron fraction distribution versus velocity in the outflows.}
    \label{fig:fig_hist}
  \end{figure}

  In Figure \ref{fig:fig_hist} we show the distributions of electron fraction measured along the wind trajectories. These trajectories were computed in a 3D numerical simulation, and are reaching the outer boundary located around 1,000 gravitational radii form the center and escape with velocities of $0.1-0.4 c$. The simulation was run over 20,000 gravitational times, which for the adopted black hole mass of 8.2 $M_{\odot}$ is equal to about 0.9 sec. The lowest values of electron fraction were carried by fast tracers launched early from the innermost regions of the disk and its surface, and reaching outer boundary at times $0.2-0.4$ seconds. 
  The $Y_{e}$ below 0.2 enabled subsequent formation of heavy elements via nucleosynthesis. That was computed using nuclear reaction network.
  %calculations.
  %a posteriori. 

\begin{figure}
  \centering
    \includegraphics[width=0.9\textwidth]{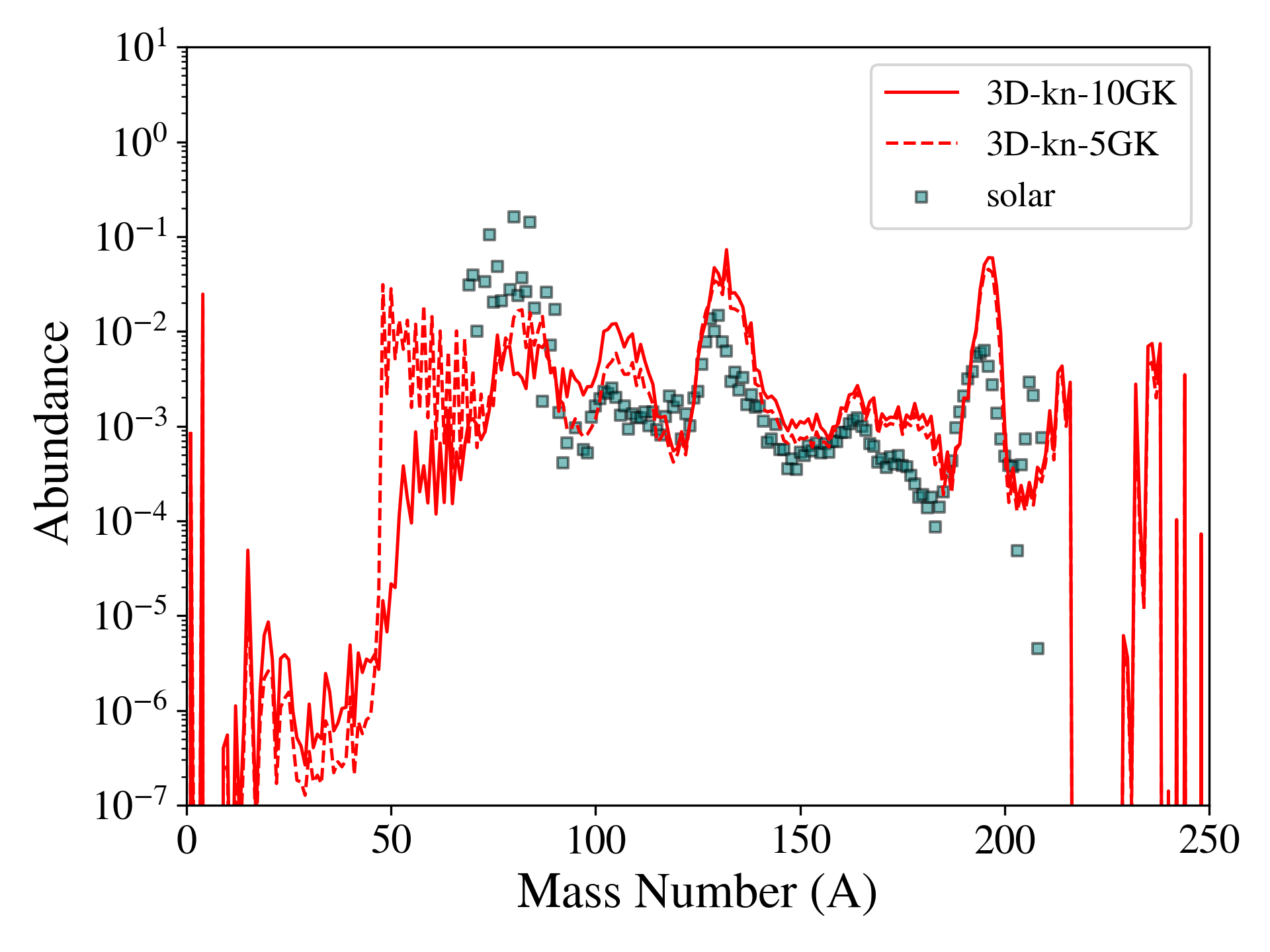}
    \caption{Abundance pattern from r-process nucleosynthesis, averaged over tracers following disk wind outflows.% launched from the BH accretion disk. 
    The model 3D-kn represents an engine composed of a 8.2 $M_{\odot}$ black hole, mildly rotating with spin $a=0.8$ and surrounded by a disk with initial mass of $0.3 M_{\odot}$. Solid and dashed lines show results from nuclear reaction network runs where the onset of r-process %network 
    after the Nuclear Statistical Equilibrium was either at $T=10 GK$ or at $T=5 GK$.}
    \label{fig:fig_abund}
  \end{figure}

  In Figure \ref{fig:fig_abund} we present the abundance pattern of isotopes synthesized via r-process in the accretion disk wind. The thick red line represents averaged abundance, with 3 characteristic peaks and an excess around mass number $A\sim 150$ due to Lanthanide production. Solar system abundance pattern is overplotted for reference. The total mass loss rate in this simulation was $9.8 M_{\odot}$ and seems sufficient to provide amounts of unstable radioactive isotopes whose decays are responsible for the kilonova emission.
  The fraction of Lanthanides produced in this scenario was about 0.0265, so that the component of the observed kilonova due to disk wind would be presumably a red (or 'purple') one, such as that observed in GRB 211211A \cite{Rastinejad2022}.
  
\section{Interaction between jet and disk wind}
\label{sect:winds}

  \begin{figure}
  \centering
    \includegraphics[width=0.5\textwidth]{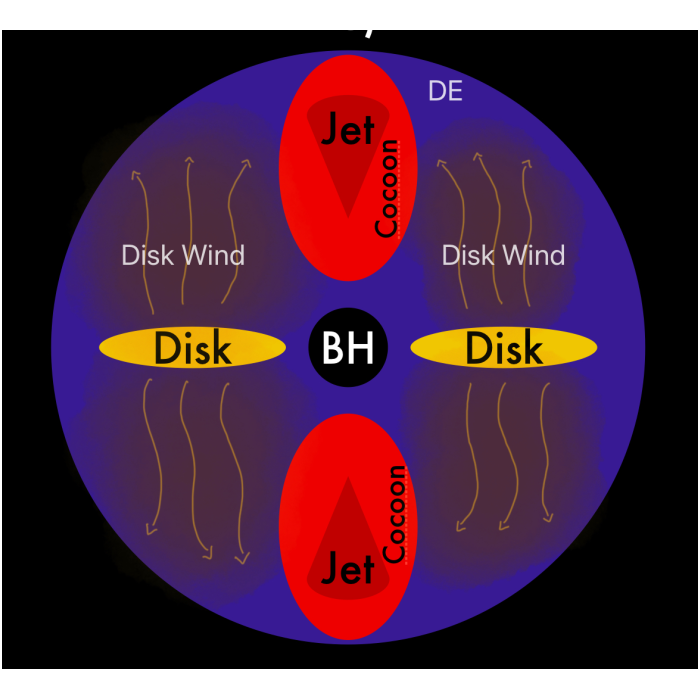}
    \caption{Schematic cartoon of a post-merger (NSNS or BHNS) system.
The ejecta are composed of the dynamical ejecta (DE) launched before the merger or soon after the HMNS was formed, and the
disk wind, formed after accreting black hole starts powering the GRB central engine. The DE undergo initial expansion into the circum-burst environment where
the jet and disk wind are propagating. In this system, the DE, disk wind, and jet are interacting with each other.}
    \label{fig:fig1}
  \end{figure}

In Figure \ref{fig:fig1}
 we present a schematic cartoon of the post-merger system, where various components are considered: central black hole formed after HMNS collapse, accretion disk, jet, its cocoon, and expanding dynamical ejecta (DE). 
 All these media are interacting with each other, as explored in detail by \cite{gerardo2025}.
The DE were adopted as spherically distributed medium, with density given by
\begin{equation}
   \rho_{\rm DE-1} = \frac{ \dot{M}}{4\pi r_{\rm inj}^2 v_{\rm ej} }\, ,
\label{eqn:dyn_floor} 
\end{equation}
or by
\begin{equation}
\rho_{\rm DE-2} (r) = \rho_{\rm DE-1} e^{-r/r_{\rm inj}}, 
\label{eqn:sph_floor}    
\end{equation}
where the exponential cutoff is taken following \cite{lazzati2021}.
The numerical values of parameters in the above relations are taken to be $\dot M = 0.01 M_{\odot} s^{-1}$ and $v_{\rm ej}=0.1$~c, for the mass loss rate and expansion velocity, respectively. The injection radius is $r_{\rm inj}=3\times 10^{8}$~cm and constitutes an inner boundary of the computational grid, to which the jet is injected.

The disk wind, on the other hand, is not spherical and has a very inhomogeneous structure. In simulations presented by \cite{gerardo2025}, the winds outflowing from post-merger accretion disk were originally modeled by two-dimensional General Relativistic MHD simulations \cite{fatemeh2023}.

As found by \cite{gerardo2025}, both DE and disk winds impact the jet propagation, and have effect on its collimation. 
In particular models where we assumed same initial jet opening angle of $15^{\circ}$, the final jet we collimated down to about $5.7^{\circ}$ or $3.6^{\circ}$, if only the disk wind, or disk wind plus DE, were interacting with this jet. This results stand for our scenario of NS-NS merger
(as adopted with particular choice od disk-to-BH mass ratio, which is typically twice larger for BH-NS than for NS-NS). For the BH-NS merger scenario, the collimated jets were not as narrow, and their final opening angles were either $12^{\circ}$ or   $13^{\circ}.2$.
The jet energy was important 
for probing the DE effect on collimation. If the jet was injected to the spherical medium only, without disk wind, it was actually collimated by the DE only for the more energetic case, with $L_{\rm jet}=2.1\times 10^{50}$~erg s$^{-1}$, however the collimation to only about $13^{\circ}$ was obtained. For weaker jet of $L_{\rm jet}=1.4 \times 10^{50}$~erg s$^{-1}$, the final opening angle was larger and jets spread to above $16^{\circ}$.

 \section{Jets from self-gravitating collapsars}
 \label{sect:collapsar}

 Most recently, we performed 3D General Relativistic MHD simulations of self-gravitating collapsars. The main aspect was to identify the conditions for jet launching and propagation in the collapsar with initially imposed strong magnetic fields, and to probe how self-gravity affects the jet when these conditions are satisfied.
 %otherwise similar.
    
\begin{figure}
    \includegraphics[width=0.5\textwidth]
    {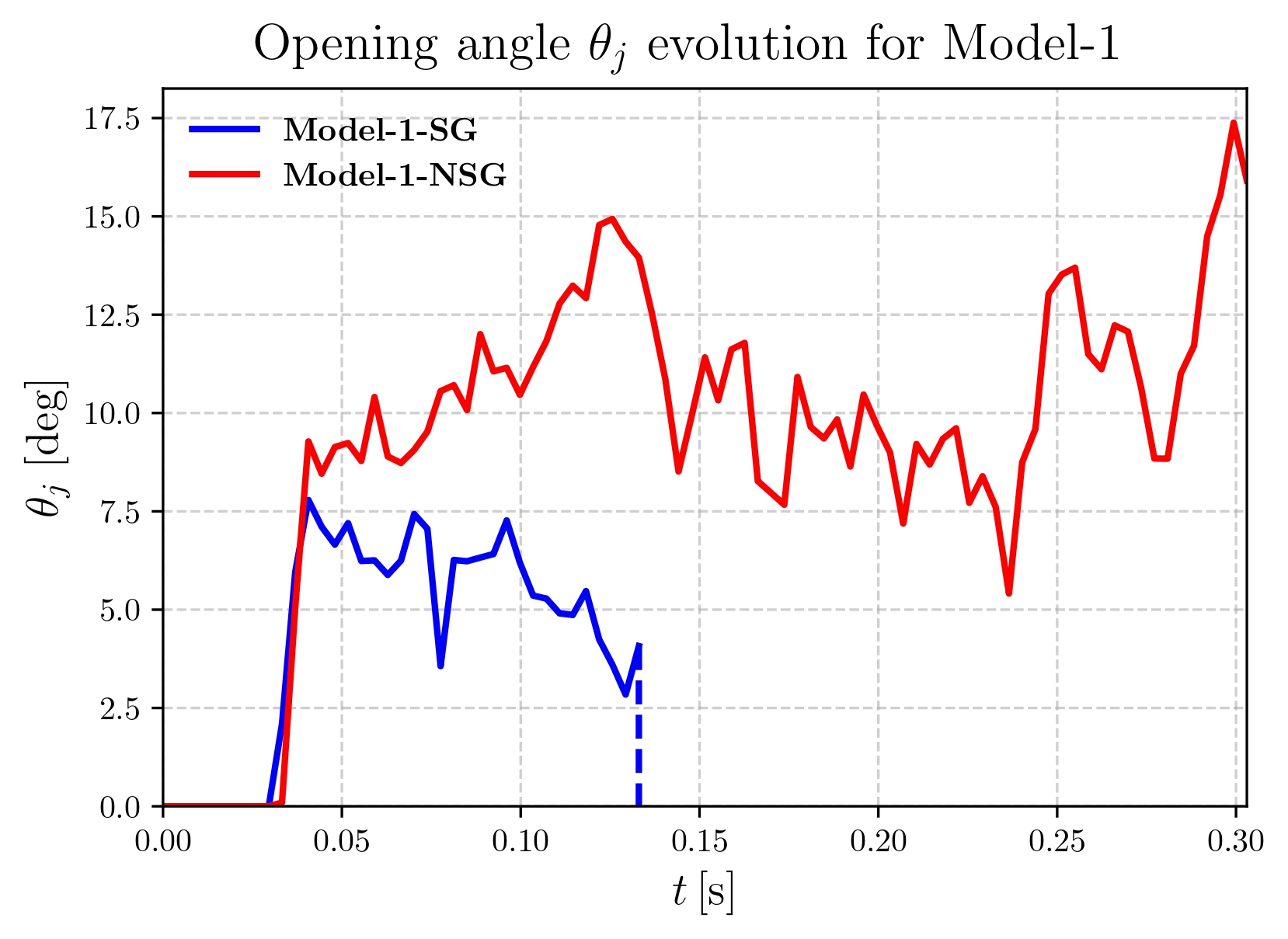}
     \includegraphics[width=0.5\textwidth]{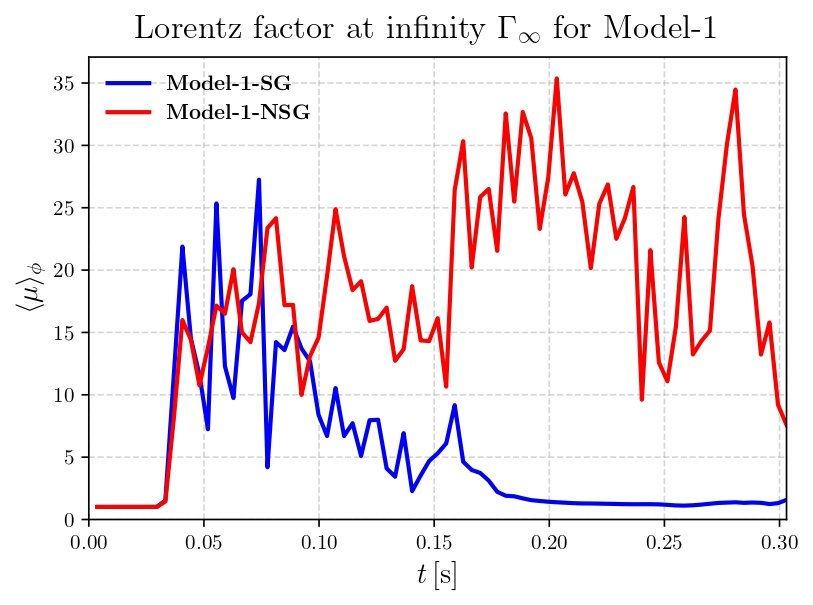}
    \caption{Jet opening angle (left) and terminal Lorentz factor (right) for the simulations of magnetized collapsars. Self-gravitating model is represented by blue lines, and the non self-gravitating case is shown with red lines.}
    \label{fig:fig2}
  \end{figure}

 In Figure \ref{fig:fig2} we show the jet opening angle and its terminal Lorentz factor, as a function of time, for two models. They have the same initial setups: slowly rotating, quasi-spherical cloud of gas, representing pre-collapse stellar core, endowed with low angular momentum (twice that required to circularize matter at the ISCO) and a uniform magnetic field. The details of this setup have been already presented in \cite{krol2021}, however that work did not include self-gravity. Now, one of the models is still non-self-gravitating (i.e. it assumed that mass of the stellar core, even though larger than the central black hole mass, is diluted and has no effect on the matter orbiting around it). The results for this model are plotted with red lines.
 The other model is the self-gravitating case, %where the self-force is 
 computed numerically in the dynamical simulation, %and is considered as a 
 where a perturbative term is added to the potential of the black hole, see \cite{janiuk2023}.

 We found, that the jet launching is in general more difficult under self-gravity. Some models exhibited a successful jet launching, and large Lorentz factors have been reached, just after the transient time when sufficiently large magnetic flux was built (MAD state) on the BH horizon. However, only in the non-SG cases we were able to observe uninterrupted jet propagation. The jets produced in SG models were narrower, and less energetic already at the beginning of the propagation. Eventually, they become quenched. 
 Our hypothesis is that the increased ram pressure of the gas subject to additional force due to self-gravity may quench the jet propagation. It can be observable as a quiescent period in the GRB prompt emission, and/or a plateau phase (Plonka \& Janiuk, 2025, submitted).

 %\section{Summary}

{\bf Acknowledgments} 
We were partially supported by grant 2023/50/A/ST9/00527 from Polish National Science Center. We also acknowledge the PL-Grid infrastructure, Cyfronet AGH, and LUMI supercomputing facility, via grants  PLG/2024/017013 and PLL/2024/07/017501 and support of the Interdisciplinary
Centre for Mathematical and Computational Modeling of the University of Warsaw under computational allocations g100-2226 and g100-2230.

\end{document}